# Signal retrieval with measurement system knowledge using variational generative model

Zheyuan Zhu[1], Yangyang Sun[1], Johnathon White[2], Zenghu Chang[2] and Shuo Pang[1]

*Abstract*— Signal retrieval from a series of indirect measurements is a common task in many imaging, metrology and characterization platforms in science and engineering. Because most of the indirect measurement processes are well-described by physical models, signal retrieval can be solved with an iterative optimization that enforces measurement consistency and prior knowledge on the signal. These iterative processes are time-consuming and only accommodate a linear measurement process and convex signal constraints. Recently, deterministic neural networks have been widely adopted to supersede iterative signal retrieval methods by approximating the inverse mapping of the measurement model. However, networks with deterministic processes have failed to distinguish signal ambiguities in ill-posed measurement systems, and the retrieved signals often lack consistency with the measurement. In this work we introduce a variational generative model to capture the distribution of all possible signals, given a particular measurement. By exploiting the known measurement model in the variational generative framework, our signal retrieval process resolves the ambiguity in the forward process, and learns to retrieve signals that satisfy the measurement with high fidelity in a variety of linear and nonlinear ill-posed systems, including ultrafast pulse retrieval, coded aperture compressive video sensing and image retrieval from Fresnel hologram.

## I. INTRODUCTION

In many areas of science and engineering, direct measurements of the signals of interest are unavailable. Ingenious measurement schemes can transform the inaccessible signals to measureable quantities, which facilitate the retrieval of the original signals. Many of such schemes, such as interferometry, tomography, and holography, have become standard measurement systems [1]–[4]. These measurement schemes, not necessarily following the dimension or sequence of original signals, further enable the reconstruction of abstract object dimensions [5]–[7], and engender more efficient acquisition processes [8]–[10].

Generally, the signal of interest, $\mathbf{f}$, needs to be retrieved from the measurement, $\mathbf{g}$. Usually, we have the knowledge of the measurement process $\mathbf{g} = A(\mathbf{f})$, where the forward operator $A(\cdot)$ describes the transformation model of the measurement system. Under this framework, the main task is to find an optimal reconstruction $\hat{\mathbf{f}}$, which can be formulated as a constrained optimization problem

$$\hat{\mathbf{f}} = \underset{\mathbf{f}}{\operatorname{argmin}} L(\mathbf{f})$$
$$= \underset{\mathbf{f}}{\operatorname{argmin}} \{\|\mathbf{g} - A(\mathbf{f})\|^2 + \lambda \phi(\mathbf{f})\}, \quad (1)$$

where $\|\mathbf{g} - A(\mathbf{f})\|^2$ describes the error between the observed measurement and the retrieved signal; the regularizer, $\phi(\mathbf{f})$, is introduced to regularize the non-uniqueness of the ill-conditioned forward model; $\lambda$ is the hyper-parameter that balances the error term and the signal regularization. The objective $L(\mathbf{f})$ can be minimized numerically with iterative algorithms [11]–[13], in which each iteration consists of two gradient descent (or proximal gradient) steps enforcing the measurement consistency and the regularization. The convergence of the iterative process requires a convex objective $L(\mathbf{f})$, which accommodates linear forward models $A(\cdot)$ and convex constraints $\phi(\cdot)$. Systems that needs to reconstruct complex-valued signal from its amplitude, such as holographic microscopy and phase retrieval, contains either non-convex forward operators or regularizers [14], and thus suffer from stagnation or failure of using iterative algorithm [15]. Moreover, the iterative process is time consuming, inadequate for many real-time applications.

The fast inference and the ability of learning a versatile mapping from measurement to signal contribute to the wide adoption of neural networks in recent years [16]–[20]. Most of these approaches use a neural network with parameters $\theta$ to approximate the inverse mapping $A_{inv}^{(\theta)}(\cdot)$. The parameters are optimized to minimize the discrepancy between ground truth $\mathbf{f}$ and inference $A_{inv}^{(\theta)}(\mathbf{g})$ based on a series of observations $\{\mathbf{f}_i\}$. Despite its simplicity and popularity, there are two major disadvantages in such neural network inversion approach. (1) The deterministic inversion cannot handle model ambiguity (i.e. one measurement corresponding to multiple possible signals), yielding reconstructions that resemble the average training instances [21]. (2) The inversion network does not use the knowledge of the measurement system, and the reconstruction is usually inconsistent with the forward model [13].

The flexibility provided by the neural networks and the knowledge of the measurement systems can be combined under the Bayesian interpretation to Eq. (1). Ref. [23]–[25] have demonstrated promising results by developing a separately trained generative model to approximate the signal prior (the regularization term), combined with the iterative algorithm for signal reconstruction. However, due to its dependency on the iterative algorithm, such approach has been demonstrated only with linear forward models, and its processing time remains too long for real-time retrieval.

[1] Zheyuan Zhu, Yangyang Sun and Shuo Pang are with CREOL, The College of Optics and Photonics, University of Central Florida. Emails: zyzhu@knights.ucf.edu, yangyang@knights.ucf.edu, pang@creol.ucf.edu
[2] Johnathon White and Zenghu Chang are with the Department of Physics and CREOL, The College of Optics and Photonics, University of Central Florida. Emails: johnathonwhite@knights.ucf.edu, Zenghu.Chang@ucf.edu
This work was supported by United States Air Force Office of Scientific Research (AFOSR) (FA9550-15-1-0037, FA9550-16-1-0013); Army Research Office (ARO) (W911NF-14-1-0383, W911NF-19-1-0224); Defense Advanced Research Projects Agency (DARPA) (D18AC00011); National Science Foundation (1806575).

In this work, we propose a signal retrieval framework based on variational generative model that allows the incorporation of the measurement system knowledge. Our model does not rely on iterative system and thus can effectively generate signal instances consistent with the forward models. In experiments, we demonstrate our approach in a variety of measurement systems, including ultrafast pulse retrieval (nonlinear problem, with phase shift ambiguity), coded aperture video compressive sensing (ill-poised linear problem), and image retrieval from Fresnel hologram (ill-poised nonlinear problem). There had not been a single framework that is applicable to all these systems and achieves similar or better reconstruction than the respective state-of-the-art methods. The paper is organized as follows. We first review the signal retrieval from Bayesian perspective in Section 2. Then we develop our variational generative model in Section 3. The experiments and the results are described in Section 4 and 5, respectively. Section 6 concludes the paper.

## II. Preliminary: Bayesian Interpretation of Signal Retrieval

From Bayesian probabilistic perspective [26], the retrieved signal $\hat{\mathbf{f}}$ should be the one that maximizes the (logarithm) posterior likelihood (maximum-a-posteriori, MAP), given the measurement $\mathbf{g}$.

$$\begin{aligned}\hat{\mathbf{f}} &= \underset{\mathbf{f}}{\mathrm{argmax}} \log p(\mathbf{f}|\mathbf{g}) \\ &= \underset{\mathbf{f}}{\mathrm{argmax}} \log \frac{p(\mathbf{g}|\mathbf{f})p(\mathbf{f})}{p(\mathbf{g})} \\ &= \underset{\mathbf{f}}{\mathrm{argmin}}(-\log p(\mathbf{g}|\mathbf{f}) - \log p(\mathbf{f})),\end{aligned} \quad (2)$$

where $p(\mathbf{g}|\mathbf{f})$ is the likelihood of observing measurement $\mathbf{g}$ from signal $\mathbf{f}$, which is determined by both the forward process $A(\cdot)$ and the noise model $p_{noise}(\mathbf{g}|A(\mathbf{f}))$ of the detection. If Gaussian noise is assumed on the detector $p_{noise}(\mathbf{g}|A(\mathbf{f})) \sim \mathcal{N}(A(\mathbf{f}), \alpha \mathbf{I})$,

$$p(\mathbf{g}|\mathbf{f}) = C_\alpha \exp\left(-\frac{\|\mathbf{g}-A(\mathbf{f})\|^2}{\alpha}\right), \quad (3)$$

where $\alpha$ is the variance that reflects the Gaussian noise level of the detector, and $C_\alpha$ is the normalization factor. The negative logarithm of $p(\mathbf{g}|\mathbf{f})$ becomes a mean squared error (MSE) of the measurement $\|\mathbf{g}-A(\mathbf{f})\|^2$ in constrained optimization (Eq. (1)). Notice that $p(\mathbf{g}|\mathbf{f})$ can also be tailored to other detector noise models such as Poisson and Binomial etc.[12], [27]–[29]. $p(\mathbf{f})$ is the prior distribution of all plausible signals, $\mathbf{f}$. If we assume $\mathbf{f}$ follows the distribution in Eq. (4),

$$p(\mathbf{f}) = C_{\beta,\Phi} \exp\left(-\frac{\|\Phi(\mathbf{f})\|_p^2}{\beta}\right), \quad (4)$$

then the signal regularizer $\phi(\mathbf{f})$ in Eq. (1) can be conceived as the negative logarithm of the prior distribution $p(\mathbf{f})$. Here the variance $\beta$ determines the regularization strength, and $C_{\beta,\Phi}$ is the normalization factor. The operator $\Phi$ transforms $\mathbf{f}$ onto the domain $\mathbf{u} = \Phi(\mathbf{f})$, where the signal representation, $\mathbf{u}$, belongs to a simple Gaussian distribution $\mathcal{N}(\mathbf{0}, \beta \mathbf{I})$, as suggested by Eq. (4). For compressed sensing settings based on sparsity ($l^1$-norm), $\Phi$ represents the projection onto domains such as wavelet [30] or total-variation [31]. With the Gaussian distribution assumptions in Eq. (3) and (4), maximizing the posterior likelihood $p(\mathbf{f}|\mathbf{g})$ reduces to the constrained optimization problem of Eq. (1).

From the Bayesian perspective, using a generative model approach to derive a more accurate prior distribution than Eq.(4) becomes a logical follow-up [23]–[25]. The prior $p(\mathbf{f})$ was trained separately from the forward model based on a series of observations $\{\mathbf{f}_i\}_{i=1}^N$. Though promising retrieval results has been demonstrated, the optimization remains a lengthy iterative process. In the next section, we will describe a framework based on conditional variational generative model to directly capture the posterior distribution, $p_\theta(\mathbf{f}|\mathbf{g})$, for solving various signal retrieval problems.

## III. Theory

Our approach implements the system forward process in the model, yet does not require conventional iterative reconstruction process. The conditional generation process handles the measurement ambiguity by introducing a latent variable $\mathbf{z}$ [20],

$$p_\theta(\mathbf{f}|\mathbf{g}) = \int p_\theta(\mathbf{f}|\mathbf{z},\mathbf{g}) p_\theta(\mathbf{z}|\mathbf{g}) d\mathbf{z}. \quad (5)$$

During the signal retrieval process, the latent variable $\mathbf{z}$ was sampled from the conditional prior $p_\theta(\mathbf{z}|\mathbf{g})$ given measurement $\mathbf{g}$, and the retrieved signal $\mathbf{f}$ is generated from the conditional variational distribution $p_\theta(\mathbf{f}|\mathbf{z},\mathbf{g})$. Both $p_\theta(\mathbf{z}|\mathbf{g})$ and $p_\theta(\mathbf{f}|\mathbf{z},\mathbf{g})$ distributions can be implemented with neural networks with parameter $\theta$.

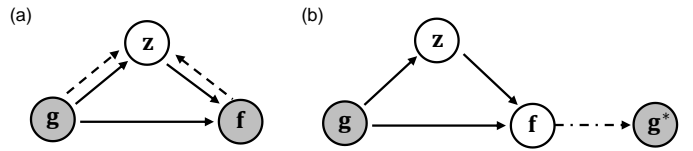

Fig. 1: Directed graphical model (solid lines) of our proposed signal retrieval network. The signal retrieval process is parameterized by $\theta$. Training of the parameters $\theta$ is assisted by introducing (a) variational inference process $q_\phi(\mathbf{z}|\mathbf{f},\mathbf{g})$ (dashed lines), (b) the known physical model $A(\cdot)$ of the measurement process (dot-dashed line). Variables in gray contain observable data in their respective models.

### A. Conditional variational inference

The objective function of the signal retrieval model is the conditional log-likelihood $\log p_\theta(\mathbf{f}|\mathbf{g}) = \sum_{i=1}^N \log p_\theta(\mathbf{f}_i|\mathbf{g}_i)$ of the observations $\{(\mathbf{f}_i, \mathbf{g}_i), i = 1, ..., N\}$ with parameters $\theta$. Due to the intractable posteriors of generative models, direct

parameter estimation is generally unfeasible. However, by substituting the objective function with its variational lower bound, the parameters can be efficiently trained [32], [33]. Through the introduction of a recognition distribution $q_\phi(\mathbf{z}|\mathbf{f}_i, \mathbf{g}_i)$ (dashed lines in Fig. 1(a)) as an approximation of the true posterior distribution $p_\theta(\mathbf{z}|\mathbf{f}_i, \mathbf{g}_i)$, the variational lower bound, $\mathcal{L}(\theta, \phi; \mathbf{f}_i, \mathbf{g}_i)$, can be derived as [34]

$$\begin{aligned}
\log p_\theta(\mathbf{f}_i|\mathbf{g}_i) &= E_{q_\phi(\mathbf{z}|\mathbf{f}_i,\mathbf{g}_i)} \log \frac{p_\theta(\mathbf{f}_i, \mathbf{z}|\mathbf{g}_i)}{q_\phi(\mathbf{z}|\mathbf{f}_i, \mathbf{g}_i)} \\
&+ KL\left(q_\phi(\mathbf{z}|\mathbf{f}_i, \mathbf{g}_i) || p_\theta(\mathbf{z}|\mathbf{f}_i, \mathbf{g}_i)\right) \\
&\geq E_{q_\phi(\mathbf{z}|\mathbf{f}_i,\mathbf{g}_i)} \log \frac{p_\theta(\mathbf{f}_i, \mathbf{z}|\mathbf{g}_i)}{q_\phi(\mathbf{z}|\mathbf{f}_i, \mathbf{g}_i)} \coloneqq \mathcal{L}(\theta, \phi; \mathbf{f}_i, \mathbf{g}_i),
\end{aligned} \quad (6)$$

where the inequality holds, because the Kullback–Leibler (KL) divergence term is always non-negative. Following the variational Bayesian approach [33], the likelihood lower bound of the inference model $\mathcal{L}$ can be expanded into

$$\begin{aligned}
\mathcal{L}(\phi, \theta; \mathbf{f}_i, \mathbf{g}_i) &= \int q_\phi(\mathbf{z}|\mathbf{f}_i, \mathbf{g}_i) \left(\log \frac{p_\theta(\mathbf{f}_i, \mathbf{z}|\mathbf{g}_i)}{q_\phi(\mathbf{z}|\mathbf{f}_i, \mathbf{g}_i)}\right) d\mathbf{z} \\
&= \int q_\phi(\mathbf{z}|\mathbf{f}_i, \mathbf{g}_i) \left(\log \frac{p_\theta(\mathbf{z}|\mathbf{g}_i)}{q_\phi(\mathbf{z}|\mathbf{f}_i, \mathbf{g}_i)} \right. \\
&\left. + \log p_\theta(\mathbf{f}_i|\mathbf{z}, \mathbf{g}_i)\right) d\mathbf{z} \\
&= -KL\left(q_\phi(\mathbf{z}|\mathbf{f}_i, \mathbf{g}_i) || p_\theta(\mathbf{z}|\mathbf{g}_i)\right) \\
&+ E_{q_\phi(\mathbf{z}|\mathbf{f}_i,\mathbf{g}_i)}(\log p_\theta(\mathbf{f}_i|\mathbf{z}, \mathbf{g}_i)) .
\end{aligned} \quad (7)$$

Here we assume the conditional prior $p_\theta(\mathbf{z}|\mathbf{g})$ is a Gaussian distribution,

$$p_\theta(\mathbf{z}|\mathbf{g}_i) = \mathcal{N}\left(\boldsymbol{\mu}_\mathbf{z}^{(\theta)}(\mathbf{g}_i), diag\left(\left[\boldsymbol{\sigma}_\mathbf{z}^{(\theta)}(\mathbf{g})\right]^2\right)\right),$$

where the mean $\boldsymbol{\mu}_\mathbf{z}^{(\theta)}$ and standard deviation $\boldsymbol{\sigma}_\mathbf{z}^{(\theta)}$ are implemented by neural networks. Similar assumption is applied to the recognition model,

$$q_\phi(\mathbf{z}|\mathbf{f}_i, \mathbf{g}_i) = \mathcal{N}\left(\boldsymbol{\mu}_\mathbf{z}^{(\phi)}(\mathbf{f}_i, \mathbf{g}_i), diag\left(\left[\boldsymbol{\sigma}_\mathbf{z}^{(\phi)}(\mathbf{f}_i, \mathbf{g}_i)\right]^2\right)\right).$$

The KL term in $\mathcal{L}(\phi, \theta; \mathbf{f}_i, \mathbf{g}_i)$ can then be explicitly expressed as

$$\begin{aligned}
&KL\left(q_\phi(\mathbf{z}|\mathbf{f}_i, \mathbf{g}_i) || p_\theta(\mathbf{z}|\mathbf{g}_i)\right) \\
&= \sum_{j=1}^{M}\left(\log \frac{\sigma_{ij}^{(\phi)}}{\sigma_{ij}^{(\theta)}} + \frac{\left(\mu_{ij}^{(\theta)} - \mu_{ij}^{(\phi)}\right)^2 + \sigma_{ij}^{(\theta)2}}{2\sigma_{ij}^{(\phi)2}} - \frac{1}{2}\right),
\end{aligned} \quad (8)$$

where $j$ is the index of elements in the $M$-dimensional vectors $\boldsymbol{\mu}_\mathbf{z}^{(\phi)}(\mathbf{f}_i, \mathbf{g}_i)$ and $\boldsymbol{\sigma}_\mathbf{z}^{(\phi)}(\mathbf{f}_i, \mathbf{g}_i)$, and their $j$-th elements are denoted as $\mu_{ij}^{(\phi)}$ and $\sigma_{ij}^{(\phi)}$. Similar notations are applied to $\boldsymbol{\mu}_\mathbf{z}^{(\theta)}(\mathbf{g}_i), \boldsymbol{\sigma}_\mathbf{z}^{(\theta)}(\mathbf{g}_i)$ as well. We also model $p_\theta(\mathbf{f}_i|\mathbf{z}, \mathbf{g}_i)$ as a Gaussian distribution, $p_\theta(\mathbf{f}_i|\mathbf{z}, \mathbf{g}_i) = \mathcal{N}(\mathbf{f}_i; \boldsymbol{\mu}_\mathbf{f}^{(\theta)}(\mathbf{z}, \mathbf{g}_i), \beta\mathbf{I})$. The second term of the lower bound becomes

$$\begin{aligned}
&E_{q_\phi(\mathbf{z}|\mathbf{f}_n,\mathbf{g}_n)}(\log p_\theta(\mathbf{f}_n|\mathbf{z}, \mathbf{g}_n)) \\
&\approx -\frac{1}{\beta L}\sum_{l=1}^{L}\left(\mathbf{f}_n - \boldsymbol{\mu}_\mathbf{f}^{(\theta)}(\mathbf{z}_l, \mathbf{g}_n)\right)^2,
\end{aligned} \quad (9)$$

where we have approximated the expectation $E_{q_\phi(\mathbf{z}|\mathbf{f}_i,\mathbf{g}_i)}$ by sampling $L$ instances of $\mathbf{z}$ from the recognition distribution $q_\phi(\mathbf{z}|\mathbf{f}_i, \mathbf{g}_i)$ as $\{\mathbf{z}_l: l = 1, ..., L\}$.

### B. Signal retrieval with measurement consistency

During the training phase, the variational inference model draws samples $\mathbf{z}$ from the recognition distribution $q_\phi(\mathbf{z}|\mathbf{f}_i, \mathbf{g}_i)$. The signal retrieval model, however, draws $\mathbf{z}$ from the conditional prior distribution $p_\theta(\mathbf{z}|\mathbf{g}_i)$. This inconsistency between the recognition distribution and conditional prior distribution was also recognized in Ref.[32]. When using the variational lower bound as the objective function, relying only on closing the KL-divergence between $q_\phi(\mathbf{z}|\mathbf{f}_i, \mathbf{g}_i)$ and $p_\theta(\mathbf{z}|\mathbf{g}_i)$ cannot provide effective training to the conditional prior. Here we take the measurement process into account and construct an alternative model to assist the training.

For the signal retrieval process, latent variable samples drawn from $p_\theta(\mathbf{z}|\mathbf{g}_i)$ capture the variance of all signals $\mathbf{f}$ that produce measurement $\mathbf{g}_i$. Naively replacing $q_\phi(\mathbf{z}|\mathbf{f}_i, \mathbf{g}_i)$ with $p_\theta(\mathbf{z}|\mathbf{g}_i)$ in the log-likelihood lower bound (Eq. (7)), in attempt to keep $\mathbf{z}$ distribution consistency, amounts to comparing all possible generated signal $\mathbf{f} = \boldsymbol{\mu}_\mathbf{f}^{(\theta)}(\mathbf{z}_l, \mathbf{g}_i)$ with a single observation $\mathbf{f}_i$. To resolve this issue, we introduce the measurement process (dot-dashed line in Fig. 1(b)) to the signal retrieval process. The expected measurement, $\mathbf{g}^*$, is generated from $\boldsymbol{\mu}_\mathbf{f}^{(\theta)}(\mathbf{z}, \mathbf{g}_i)$ via the forward process $\mathbf{g}^* = A(\boldsymbol{\mu}_\mathbf{f}^{(\theta)}(\mathbf{z}, \mathbf{g}_i))$. For all the $\mathbf{z}$ sampled from $p_\theta(\mathbf{z}|\mathbf{g}_i)$, we maximize the likelihood of generating the expected measurement $\mathbf{g}^*$ given point $\mathbf{g}_i$, as defined by the detection model. Applying Jensen's inequality, a lower bound of this likelihood $\mathcal{L}_r(\theta; \mathbf{g}^*, \mathbf{g}_i)$ can be derived and used as the objective function of the retrieval process.

$$\begin{aligned}
\log p(\mathbf{g}^*|\mathbf{g}_i) &= \log \int p_\theta(\mathbf{g}^*|\mathbf{z}, \mathbf{g}_i) p_\theta(\mathbf{z}|\mathbf{g}_i) \, d\mathbf{z} \\
&\geq \int \log p_\theta(\mathbf{g}^*|\mathbf{z}, \mathbf{g}_i) \, p_\theta(\mathbf{z}|\mathbf{g}_i) d\mathbf{z} \\
&= E_{p_\theta(\mathbf{z}|\mathbf{g}_i)}(\log p_\theta(\mathbf{g}^*|\mathbf{z}, \mathbf{g}_i)) \\
&\approx -\frac{1}{\alpha L}\sum_{l=1}^{L}\left(A\left(\boldsymbol{\mu}_\mathbf{f}^{(\theta)}(\mathbf{z}_l, \mathbf{g}_i)\right) - \mathbf{g}_i\right)^2 \\
&\coloneqq \mathcal{L}_r(\theta; \mathbf{g}^*, \mathbf{g}_i),
\end{aligned} \quad (10)$$

where we have assumed Gaussian noise model on the detector $\mathbf{g}^* \sim \mathcal{N}(\mathbf{g}_i, \alpha\mathbf{I})$. The expectation $E_{p_\theta(\mathbf{z}|\mathbf{g}_i)}$ in Eq. (10) is approximated by sampling $L$ instances of from the conditional prior distribution $p_\theta(\mathbf{z}|\mathbf{g}_i)$ as $\{\mathbf{z}_l: l = 1, ..., L\}$. By adding in the measurement processes, we essentially construct a variational autoencoder for measurement $\mathbf{g}$, and the objective function promotes forward model consistency. We jointly train the retrieval model alongside the inference model with a hybrid objective function [32],

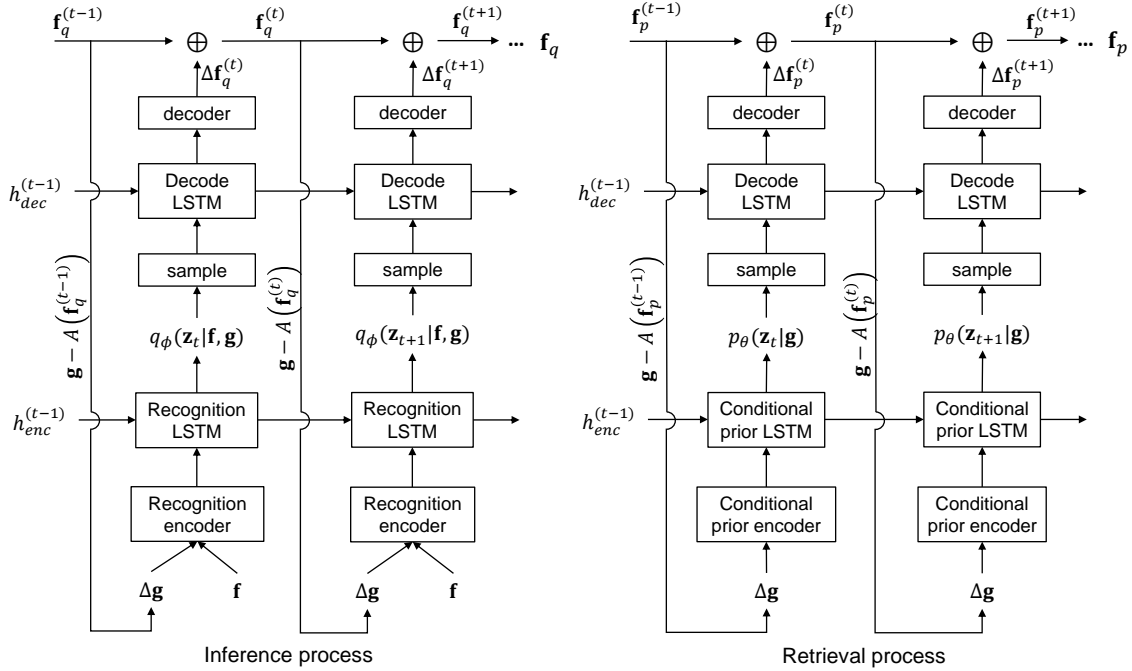

Fig. 2: Recurrent structure of the conditional variational generative network at time stamp $t$. Each box represents a network and arrows represent data flow. In both models, the generated $\Delta \hat{\mathbf{f}}$ from all previous time stamps are aggregated to obtain $\mathbf{f}$ at $t$. The weights of the decoder are shared between inference and retrieval process.

$$\mathcal{L}_h(\phi, \theta, \mathbf{f}_i, \mathbf{g}_i) = \gamma \mathcal{L}(\phi, \theta; \mathbf{f}_i, \mathbf{g}_i) + (1-\gamma)\mathcal{L}_r(\theta; \mathbf{g}^*, \mathbf{g}_i),$$
(11)

where the hyperparameter $\gamma$ balances the weight between the two models.

## IV. EXPERIMENTS

The conditional variational generative model consisted of inference and retrieval processes. The implementations of both processes are detailed in Fig. 2. The encoder of the inference process takes in $\mathbf{f}$ and $\mathbf{g}$ as inputs. The encoder of the retrieval process only accepts one input, $\mathbf{g}$. The mapping from latent domain to the signal domain $\mathbf{F}$ is performed by a decoder whose weights are shared in both processes. Inspired by the conventional iterative algorithms, our network adopted a recurrent construction [35] with LSTM cells. Specifically, outputs from the both processes, $\mathbf{f}_q$ and $\mathbf{f}_p$, are initialized as 0. Each recurrence generates an increment $\Delta \mathbf{f}^{(t)}$ to $\mathbf{f}$ from the discrepancy $\mathbf{g} - A(\mathbf{f}^{(t-1)})$ between observed measurement $\mathbf{g}$ and the previous estimate $A(\mathbf{f}^{(t-1)})$. For the first recurrence, this discrepancy was set to $\mathbf{g}$. During the retrieval process, one measurement $\mathbf{g}_i$ from the test dataset and one sample $\mathbf{z}_l \sim p_\theta(\mathbf{z}|\mathbf{g}_i)$ are fed into the generative network $p_\theta(\mathbf{f}|\mathbf{g}_i, \mathbf{z}_l)$ to obtain one reconstruction instance $\hat{\mathbf{f}}_i$. As a comparison, we also trained a single-pass deterministic neural network based on the structure of the retrieval process. The sampling process and physical model were removed in the deterministic network. The loss function of the deterministic network consisted only of the MSE on $\mathbf{f}$. All the networks and physical model of the measurement process were implemented in TensorFlow 1.9.0 and Python 3.6 environment.

The reconstruction performance is evaluated by peak signal-to-noise ratio (PSNR), which is defined as PSNR= $10 \times \log_{10}(\max(\mathbf{f}_i)/\text{MSE})$, where MSE=$\frac{1}{\dim(\mathbf{f}_i)}\|\hat{\mathbf{f}}_i - \mathbf{f}_i\|^2$ is the mean square error between the ground truth $\mathbf{f}_i$ and the reconstructed instance $\hat{\mathbf{f}}_i$ across all dimensions of $\mathbf{f}$. The fidelity, defined as the PSNR between the measurements generated from reconstruction $\mathbf{g}^* = A(\hat{\mathbf{f}}_i)$ and ground truth $\mathbf{g}_i$, quantifies how well the reconstructions match the physical model of the measurement process. We compared the reconstruction performance between our model and deterministic networks with three signal retrieval examples, detailed in the following subsections.

### A. Ultrafast pulse retrieval

The forward process of ultrafast pulse retrieval was established based on the theory in Ref.[19]. The streak trace is a series of photoelectron spectra $I(K, \tau)$ arising from the interaction between an attosecond extreme ultraviolet (XUV) pulse $\tilde{\vec{E}}_{XUV}(t)$ and a femtosecond infrared (IR) dressing field $\tilde{\vec{E}}_{IR}(t)$ under different time delays $\tau$

$$I(K,\tau) = \left| \int \tilde{\vec{E}}_{XUV}(t-\tau) \cdot \vec{d} \exp{\mathbb{i}\phi_G(K,t)} \exp\left(-\mathbb{i}\frac{K+I_p}{\hbar}t\right) dt \right|^2,$$
(12)

where $\mathbb{i}$ is the imaginary unit, $K$ is the kinetic energy of the photoelectron; $\hbar$ is the Plank's constant; $I_p$ is the ionization potential; $\vec{d}$ is the dipole transition matrix element from the ground state to the continuum state [4], and is assumed to be constant[36]; $\phi_G$ is a phase gate on the photoelectron wave $\tilde{\vec{E}}_{XUV}(t-\tau) \cdot \vec{d}$, and is determined by the IR dressing field via [36]

$$\phi_G(K,t) = -\int_t^\infty \left[\vec{v}\cdot\vec{A}(t') + \frac{\vec{A}^2(t')}{2}\right]dt', \quad (13)$$

where $\vec{v}$ is the momentum of the electron and is related to the kinetic energy via $K = v^2/2$; $\vec{A}(t) = -\partial\tilde{\vec{E}}_{IR}/\partial t$ is the vector potential of the IR dressing field. In streaking experiments, the X-ray and IR fields are linearly polarized along the same directions, such that $\tilde{\vec{E}}_{XUV}$ and $\tilde{\vec{E}}_{IR}$ could both be reduced to scalar field $\tilde{E}_{XUV}$ and $\tilde{E}_{IR}$.

The XUV and IR pulses were created by imposing the experimental XUV and IR spectra on their corresponding spectral phases $\tilde{E}(\omega) = \sqrt{S(\omega)}\exp i\phi(\omega)$, and Fourier-transformed into the time domain for streak calculation. The spectral phase term was expressed as a 5th order polynomial function $\phi(\omega) = \sum_{i=0}^{5} k_i\omega^i$. The number of sampling for XUV and IR spectra $\tilde{\mathbf{E}}_{XUV}$ and $\tilde{\mathbf{E}}_{IR}$ was 200 and 20, respectively. The input of this forward model, $\mathbf{f} \in \mathbb{R}^{440}$, was a concatenated vector representing the real and imaginary part of the XUV and IR spectra,

$[\text{Re}(\tilde{\mathbf{E}}_{XUV}), \text{Im}(\tilde{\mathbf{E}}_{XUV}), \text{Re}(\tilde{\mathbf{E}}_{IR}), \text{Im}(\tilde{\mathbf{E}}_{IR})]$. The output of the forward model, $\mathbf{g} \in \mathbb{R}^{256\times 35}$, was the discretized streak trace $I$ in terms of 256 energies $K$ ranging from 50 to 305 eV, and 35 time delays $\tau$ from -8 fs to 8 fs. Since the carrier envelope phase (CEP) term $k_0$ of the XUV pulse does not affect the streak intensity, XUV pulses with the same phase coefficients except $k_0$ would yield identical streak traces $\mathbf{g}$, creating ambiguities in the training dataset.

The network for ultrafast pulse retrieval consisted of convolutional encoders, fully-connected LSTM cells and decoder. The network was trained on streak traces simulated from 10000 XUV and IR pulses with random phase coefficients $k_0$ to $k_5$ for 100 epochs. The test dataset contained another 100 streak traces. For each streak trace, we sampled 10 reconstruction instances from the approximated posterior distribution $p_\theta(\mathbf{f}|\mathbf{g})$.

*B. Coded aperture video compressive sensing*

Video compressive sensing encodes fast-moving scenes with alternating masks on the conjugate image plane so that they can be captured by a slow camera [5]. Each low-frame-rate measurement recorded on the camera, $I(x,y)$, is the high-speed scene $f(x,y,t)$ encoded by a series of rapidly-changing mask $M(x,y,t_i)$

$$I(x,y) = \int_{t=0}^{K\tau} f(x,y,t) \sum_{i=1}^{K} M(x,y,t_i) \text{rect}(\frac{t-t_i-\tau/2}{\tau}) dt, \quad (14)$$

where $1/\tau$ is the frequency of the mask. The frame rate of the camera, $1/(K\tau)$, is $K$ times slower than the mask frequency.

In this example, we demonstrate the compression of $K=4$ color frames into 1 measurement with random binary masks. The number of pixels in both the high-speed scene and measurement were $N \times N$ ($N=64$). The spatial-encoding binary masks apply to all color channels, and are represented by a Kronecker product $\mathbf{M}_i = \mathbf{m}_i \otimes \mathbf{1}^{1\times 1\times 3}, i=1,2,3,4$, where $\mathbf{m}_i \in \{0,1\}^{N\times N}$ denotes the transmittance of the mask, and $\mathbf{1}^{1\times 1\times 3}$ is a unit tensor along the dimension of color channels. Let $\{\mathbf{f}_i \in \mathbb{R}^{N\times N\times 3}, i=1,2,3,4\}$ denote the 4 color frames from the fast-moving scene within one measurement frame. The measurement $\mathbf{g} \in \mathbb{R}^{N\times N\times 3}$ is given by $\mathbf{g} = \sum_{i=1}^{4} \mathbf{M}_i \odot \mathbf{f}_i$, where $\odot$ denotes the element-wise product between tensors. This measurement process can be described by a linear forward model $\mathbf{g} = \mathbf{A}\mathbf{f}$, in which $\mathbf{f}$ and $\mathbf{g}$ are vectorized into $\mathbb{R}^{12N^2}$ and $\mathbb{R}^{3N^2}$ respectively; $\mathbf{A}$ is concatenated from 4 diagonal matrices $[diag(\mathbf{M}_1), diag(\mathbf{M}_2), diag(\mathbf{M}_3), diag(\mathbf{M}_4)]$, where $\mathbf{M}_i$ is vectorized into $\mathbb{R}^{3N^2}$.

The network for video compressive sensing employed convolutional layers in encoders, LSTM cells and decoders. The network was trained on random four-image combinations from the ImageNet database, and tested on 100 traffic video segments in DynTex library.

*C. Image retrieval from Fresnel hologram*

We constructed an Fresnel in-line hologram forward model based on the setup in Ref.[37]. Coherent, parallel beam illumination ($\lambda$=635nm) was assumed in the forward model and the propagation distance $z$ between the object and the detector plane was set to 400mm. The intensity on the camera is the interference between the propagated field and the reference beam

$$I(x,y) = \left|A_{ref} + \tilde{E}_{prop}(x,y)\right|^2, \quad (15)$$

where $A_{ref}$ represents the parallel, on-axis reference field. The complex field, $\tilde{E}_{prop}$, is given by the Fresnel propagation of incident field $\tilde{E}_o$,

$$\tilde{E}_d(x,y) = \int \tilde{E}_o(x_0,y_0) \exp\left[\frac{i\pi}{z\lambda}((x-x_0)^2 + (y-y_0)^2)\right] dx_0 dy_0, \quad (16)$$

where $(x_0,y_0)$ is the spatial coordinates of the incident field, and $z$ is the propagation distance.

This example considers the retrieval of a real object from its in-inline Fresnel hologram. The fields on the object $\tilde{\mathbf{E}}_o \in \mathbb{C}^{64\times 64}$ and camera plane $\tilde{\mathbf{E}}_d \in \mathbb{C}^{64\times 64}$ were both discretized into 64X64 pixels, with a pixel size of 50μm. The input of the forward model, $\mathbf{f} \in \mathbb{R}^{64\times 64}$, was a zero-padded MNIST digit representing the real part of $\tilde{\mathbf{E}}_o$. The imaginary part of $\tilde{\mathbf{E}}_o$ was set to zero. Let $\mathbf{x}_0, \mathbf{y}_0$ denote the coordinates of pixels on the object plane, the complex field on the camera plane, $\tilde{\mathbf{E}}_d$, can be formulated as a two-dimensional, discrete convolution between input field $\tilde{\mathbf{E}}_o$ and a quadratic phase kernel $\tilde{\mathcal{F}} = \exp\left[\frac{i\pi}{z\lambda}(\mathbf{x}_o^2 + \mathbf{y}_o^2)\right]$. The measured intensity, $\mathbf{g}$, is the squared modulus of the complex field $\tilde{\mathbf{E}}_d$. As a result, we adopted convolutional structures in the encoders, LSTM cells and decoder. The network was trained on Fresnel holograms simulated from 10000 MNIST training digits for 40 epochs, and tested on 1000 pairs of holograms and digits from the MNIST test dataset.

## V. RESULTS AND DISCUSSIONS

### A. Ultrafast pulse retrieval

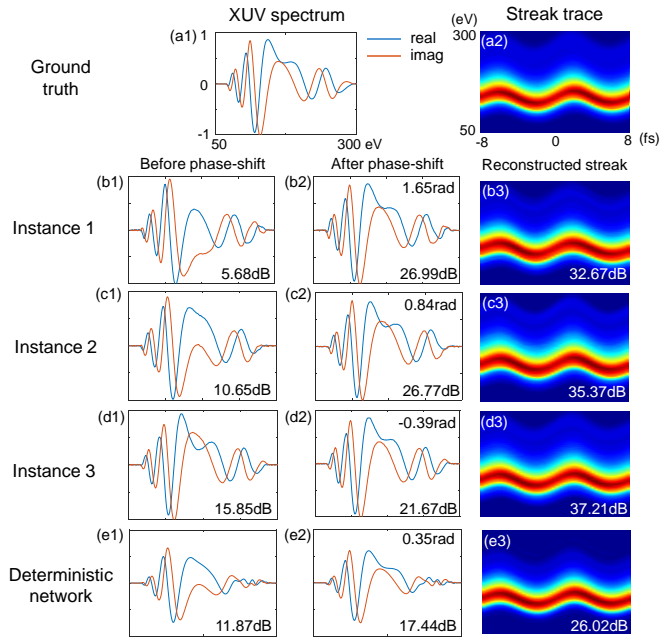

Fig. 3: Reconstructions from the ultrafast pulse retrieval experiment: (a) Ground truth of the real and imaginary part of the XUV spectrum and its simulated streak trace. The IR spectrum is not shown in the figure. (b-d) Three instances of retrieved XUV spectrum from our method (b1-d1), their phase-shifted variant (b2-d2), and the streak trace (b3-d3) calculated from each instance. (e) Retrieved XUV spectrum, its phase-shifted variant and streak trace from the deterministic network.

We first demonstrate the capability of our model to resolve ambiguities of a nonlinear forward model in the ultrafast pulse retrieval example. Fig. 3 displays the real and imaginary part of the XUV spectrums, along with their corresponding streak traces calculated from the forward process $A(\cdot)$. An XUV pulse in test dataset (Fig. 3(a1)) produced the streak trace in Fig. 3 (a2), which was fed into the trained signal retrieval process. Three instances of the retrieved XUV spectrums from Fig. 3 (a2) are shown in Fig. 3 (b1-d1), with PSNR of 5.68, 10.65 and 15.85dB, respectively, compared with the ground truth in Fig. 3 (a). Yet their high measurement fidelities suggest that these instances belong to the phase-shift ambiguities of the same streak trace. For each of the instances, we were able to shift the carrier envelop phase $k_0$ by the average phase difference within 100~300 eV, and match the retrieved XUV spectrum and ground truth with good consistency. The amount of phase-shift was 1.65, 0.84 and -0.39 radians, respectively for Fig. 3 (b2-d2), with PSNR of 26.99, 26.77 and 21.67 after the phase shift. In contrast, the deterministic network generates identical reconstructions similar to the average of the ambiguity instances. The XUV spectrum in Fig. 3 (e1) cannot be phase-shifted to match the ground truth, and exhibits low fidelity (Fig. 3 (e3)) compared with the actual measurement. Table 1 summarizes the average PSNR and fidelity of the reconstructions from the 100 test streak traces, each generating 10 instances of XUV spectrums. The high fidelity of our method indicates that it can generate different reconstruction instances satisfying the measurement forward model, a capability not possessed by deterministic network. To reach similar reconstruction fidelity from a deterministic network requires manually removing the ambiguity instances from the training data.

TABLE 1:
PSNR AND FIDELITY OF THE RECONSTRUCTED PULSE WITH DETERMINISTIC NETWORK AND OUR METHOD

|  | Deterministic network | Ours |
|---|---|---|
| PSNR | 11.08 | 13.87 |
| Fidelity | 23.10 | 31.49 |

### B. Coded aperture video compressive sensing

We trained our model on 4X coded aperture compression forward model, and tested it on compressed video frames. Fig. 4 (a) shows the ground truth of the 4 frames and the compressive measurement. The 4 reconstructed frames are shown in Fig. 4(b), along with the compressive measurement from the reconstructed frames. As a comparison, we also performed iterative maximum-a-posteriori reconstructions with TV prior and deep pixel-level prior (DPP), shown in Fig. 4 (c-d), respectively. The optimized strength of TV prior was $10^{2.0}$. The optimal step parameter of the DPP network, analogous to the regularization strength, was 0.5. The PSNR and fidelity of TV, DPP and our model are listed in Table 2. We speculate the lower fidelity is attributed to having only 3 recurrences in our model, which is currently limited by the computation power of GPU. However, it is worth noting that the reconstruction time (0.13s) of our trained model was orders of magnitude shorter than DPP, which required hundreds of iterations and took 197.3s. Both our model and DPP outperform TV thanks to their more realistic prior distributions.

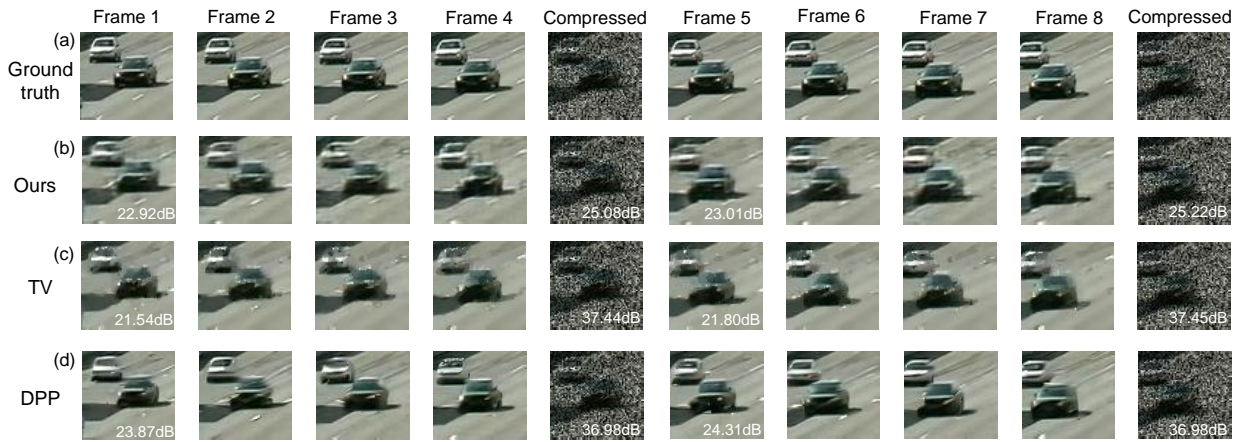

Fig. 4: Reconstructed frames from 4X video compression model: (a) Ground truth of the frames and simulated compressed measurement. (b-d) Reconstructed frames and measurements from our model, TV and Deep Pixel-level Prior (DPP), respectively. The number on the first frame indicates the PSNR of all 4 retrieved frames. The number on the compressed measurement indicates the fidelity of the measurement calculated from 4 retrieved frames.

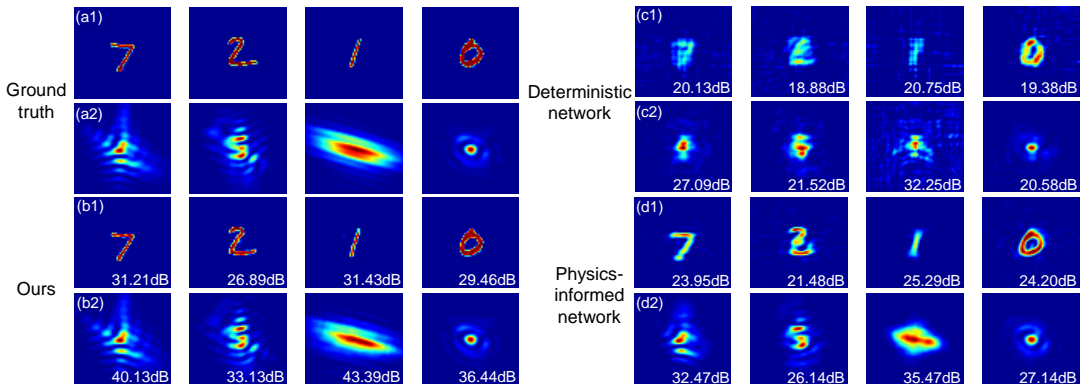

Fig. 5: Reconstructed images from Fresnel hologram: (a) Ground truth of the image and simulated Fresnel hologram intensity. (b-d) Reconstructed images and holograms from our model, deterministic network, and physics-informed network.

TABLE 2:
PSNR AND FIDELITY OF THE RECONSTRUCTIONS USING TV, DPP AND OUR METHOD

|  | TV | DPP | Ours |
|---|---|---|---|
| PSNR | 22.07 | 22.44 | 22.14 |
| Fidelity | 37.05 | 37.12 | 24.43 |

*C. Image retrieval from in-line Fresnel hologram*

In this experiment, we demonstrate the performance of our model in retrieving the image from Fresnel hologram. The Fresnel holograms (Fig. 5(a2)) simulated from the MNIST test images (Fig. 5 (a1)) were fed into the deterministic, physics-informed and our model trained on the holograms simulated from MNIST training dataset. The reconstructions were then forward propagated to the detector plane to evaluate the fidelity. As a comparison, we also performed reconstructions from a deterministic network, and a physics-informed network [37] that adds a Fresnel back-propagation operation before the deterministic network. Table 3 lists the PSNR and fidelity of all the test images retrieved from deterministic network, physics-informed network and our model with comparable structures. The fidelity of our model is better than both deterministic and physics-informed neural networks. Though physics-informed network embeds the Fresnel back-propagation as its first layer, the back-propagated image still suffers from the twin image artifact, which needs to be corrected by the subsequent deterministic neural network. In our model, we apply the Fresnel forward propagation to the intermediate reconstruction and feed the error of the measurement back into the encoder, thus achieving a higher fidelity via direct enforcement of the forward model on the reconstructed image.

TABLE 3:
PSNR AND FIDELITY OF THE RECONSTRUCTED IMAGE FROM HOLOGRAMS WITH DETERMINISTIC NETWORK, PHYSICS-INFORMED NETWORK AND OUR METHOD

|  | Deterministic network | Physics-informed network | Ours |
|---|---|---|---|
| PSNR | 19.35 | 22.83 | 27.21 |
| Fidelity | 23.46 | 28.86 | 35.30 |

## VI. CONCLUSION

We have proposed a model-based conditional generative network for solving a wide variety of signal retrieval problems, including coded aperture video compressive sensing, image retrieval from Fresnel hologram and ultrafast pulse retrieval. The proposed framework exploits the known forward process of the measurement systems to train the conditional variational generative model. Compared with deterministic neural network that approximates the inversion of the forward process, our variational generative network resolves ambiguities in the training dataset, and demonstrates high-fidelity reconstructions that are consistent with the measurement process for both linear and non-linear forward models. We envision our framework as a general signal retrieval pipeline for a variety of measurement processes in which the indirect measurement obeys a physical model.

### DATA AND CODE AVAILABILITY

The implementation of the proposed model will be made available online. The forward process of the ultrafast pulse retrieval example is available from the corresponding author upon reasonable request.